

\input harvmac

%
%



\hoffset -.24in
\hsize    5.5in
\baselineskip=20pt plus 2pt minus 1pt

%
%
%
%

\font\csc  =cmcsc10
\font\bigrm=cmb10 scaled \magstep1


\def\ha{{1 \over 2}}

\def\->{\rightarrow}     \def\<-{\leftarrow}
\def\<{\langle}          \def\>{\rangle}
\def\[{\left [}          \def\]{\right ]}
\def\({\left (}          \def\){\right )}
\def\|{\vert}

\def\Pmat#1{\left(\matrix{#1}\right)}
\def\Smat#1{\left[\matrix{#1}\right]}

\def\apar{\noalign{\vskip 2mm}}


\def\bZ{{\bf Z}} \def\bC{{\bf C}} \def\iO{{\it O}}


\def\a{\alpha}
\def\b{\beta}
\def\d{\delta}	    \def\D{\Delta}
\def\e{\epsilon}
\def\g{\gamma}	    \def\G{\Gamma}
\def\k{\kappa}
\def\la{\lambda}

\def\r{\rho}
    
	    \def\T{\Theta}    \def\vt{\vartheta}


\def\M{\overline M}    \def\tR{\tilde R}
\def\tPsi{\tilde\Psi}  \def\tPhi{\tilde\Phi}
\def\cR{\check R}      \def\cC{\check C}

\def\VO{$q$-vertex operator}
\def\VOs{$q$-vertex operators}
\def\Uqsl2{$ U_q(\widehat{sl_2}) $}

\def\opeOO{ \<j_4\|\,\Phi_{\ha,}{}^{  0}_{  0}(z_3)
                   \,\Phi_{j_2,}{}^{M  }_{M  }(z_2)\,\|j_1\> }
\def\opeOI{ \<j_4\|\,\Phi_{\ha,}{}^{  0}_{  1}(z_3)
                   \,\Phi_{j_2,}{}^{M  }_{M-1}(z_2)\,\|j_1\> }
\def\opeIO{ \<j_4\|\,\Phi_{\ha,}{}^{  1}_{  0}(z_3)
                   \,\Phi_{j_2,}{}^{M-1}_{M  }(z_2)\,\|j_1\> }
\def\opeII{ \<j_4\|\,\Phi_{\ha,}{}^{  1}_{  1}(z_3)
                   \,\Phi_{j_2,}{}^{M-1}_{M-1}(z_2)\,\|j_1\> }

\def\opeOOr{\<j_4\|\,\Phi_{j_2,}{}^{M  }_{M  }(z_2)
                   \,\Phi_{\ha,}{}^{  0}_{  0}(z_3)\,\|j_1\> }
\def\opeOIr{\<j_4\|\,\Phi_{j_2,}{}^{M  }_{M-1}(z_2)
                   \,\Phi_{\ha,}{}^{  0}_{  1}(z_3)\,\|j_1\> }
\def\opeIOr{\<j_4\|\,\Phi_{j_2,}{}^{M-1}_{M  }(z_2)
                   \,\Phi_{\ha,}{}^{  1}_{  0}(z_3)\,\|j_1\> }
\def\opeIIr{\<j_4\|\,\Phi_{j_2,}{}^{M-1}_{M-1}(z_2)
                   \,\Phi_{\ha,}{}^{  1}_{  1}(z_3)\,\|j_1\> }


\lref\Ra{K. Aomoto,
`` A note on holonomic $q$-difference systems ''
   Algebraic Analysis I, ed. by M. Kashiwara and T. Kawai,
   Academic Press, San Diego (1988) 25-28.}

\lref\Rabf{G.E. Andrews, R.J. Baxter and P.J. Forrester,
`` Eight-vertex SOS model and generalized Rogers-Ramanujan-type identities ''
   Jour. Stat. Phys. {\bf 35} (1984) 193-266.}

\lref\Rabg{A. Abada, A.H. Bougourzi and M.A. El Gradechi,
`` Deformation of the Wakimoto construction ''
   preprint (1992).}

\lref\Rakm{K. Aomoto, Y. Kato and K. Mimachi,
`` A solution of Yang-Baxter equation as connection coefficidnts
   of a holonomic $q$-difference system ''
   Duke Math. Jour. {\bf 65} IMRN (1992) 7-15.}

\lref\Raty{H. Awata, A. Tsuchiya and Y. Yamada,
`` Integral formulas for the WZNW correlation functions ''
   Nucl. Phys. {\bf B365} (1991) 680-696.}

\lref\RayI{H. Awata and Y. Yamada,
`` Fusion rules for the fractional level $\widehat{sl(2)}$ algebra~''
   Mod. Phys. Lett. {\bf A7} (1992) 1185-1195.}

\lref\RayII{H. Awata and Y. Yamada,
`` Fusion rules for the $q$-vertex operators of $U_q(\widehat{sl_2})$~''
   Mod. Phys. Lett. {\bf A7} (1992) 2235-2243.}




\lref\Rdjmo{E. Date, M. Jimbo, T. Miwa and M. Okado,
`` Solvable lattice models ''
   RIMS Preprint 590 (1987).}

\lref\Rdjo{E. Date, M. Jimbo and M. Okado,
`` Crystal base and $q$-vertex operators ''
   Osaka Univ. Math. Sci. preprint 1 (1991).}

\lref\Rdfjmn{B. Davies, O. Foda, M. Jimbo, T. Miwa and A. Nakayashiki,
`` Diagonalization of the XXZ hamiltonian by vertex operators ''
   RIMS preprint (1992).}


\lref\Rfl{G. Felder and A. LeClair,
`` Restricted quantum affine symmetry of perturbed minimal conformal models ''
   Int. J. Mod. Phys. {\bf A7} (1992) 239-278.}

\lref\Rfr{I.B. Frenkel and N.Yu. Reshetikhin,
`` Quantum affine algebras and holonomic difference equations ''
 Commun. Math. Phys. {\bf 146} (1992) 1-60.}

\lref\Riijmnt{M. Idzumi, K. Iohara, M. Jimbo, T. Miwa, T. Nakashima
   and T. Tokihiro,
`` Quantum affine symmetry in vertex models ''
   RIMS preprint (1992).}



\lref\Rmali{F.G. Malikov,
`` Quantum groups: Singular vectors and BGG resolution ''
   RIMS Preprint 835 (1991). }

\lref\RmatI{A. Matsuo,
`` Jackson integrals of Jordan-Pochhammer typ
   and quantum Knizhnik- {\break}Zamolodchikov equations ''
   preprint (1991);
`` Quantum algebra structure of certain Jackson integrals ''
   preprint (1992).  }

\lref\RmatII{A. Matsuo,
`` Free field representation of quantum affine algebra $U_q(\widehat{sl_2})$ ''
   preprint (1992).}

\lref\Rmi{K. Mimachi,
`` Connection problem in holonomic $q$-difference system
   associated with a Jackson integral of Jordan-Pochhammer type ''
   Nagoya Math. Journ., {\bf 116} (1989) 149-161.}


\lref\Rr{N. Reshetikhin,
`` Jackson-type integrals, Bethe vectors, and solutions
   to a difference analog of the Knizhnik-Zamokodchikov system ''
   preprint (1991).}

\lref\Rsh{J. Shiraishi,
`` Free boson representation of $U_q(\widehat{sl_2})$ ''
   preprint UT-617 (1992).}

\lref\Rsm{F.A. Smirnov,
`` Dynamical symmetries of massive integrable models~ ''
   Int. J. Mod. Phys. {\bf A7} Supplement 1 (1992) 813-837;839-818.}

\lref\Rtk{A. Tsuchiya and Y. Kanie,
`` Vertex operators in conformal field theory on $ P^1$
   and monodromy representations of braid group ''
   Advanced Studies in Pure Math. {\bf 16} (1988) 297-372,
   Lett. Math. Phys. {\bf 13} (1987) 303.}

\def\Fadd{\rm
E-mail address : hawa@theory.kek.jp
}

\rightline{\vbox{\hbox{EPHOU-92-013}
                 \hbox{November 1992} }}

\vskip.8in
\centerline{\bigrm Exchange Relations for the {\VOs} of \Uqsl2 }
\vskip.4in
\centerline{{\csc Hidetoshi Awata} \foot\Fadd}
{\it
\vskip.3in
\centerline{Dept. of Physics Hokkaido University,}
\centerline{Sapporo 060, Japan}
}
\rm
\vskip .6in
\centerline{\bf Abstract}
\vskip.3in

 We consider the $q$-deformed Knizhnik-Zamolodchikov equation
for the two point function of $q$-deformed vertex operators of {\Uqsl2}.
We give explicitly the fundamental solutions,
the connection matrices and the exchange relations
for the {\VOs} of spin $1/2$ and $j \in \ha\bZ_{\geq 0}$.
Consequently, we confirm that the connection matrices are equivalent to
the elliptic Boltzman weights of IRF type
obtained by the fusion procedure from ABF models.

\vskip.8in
\leftline{hep-th/9211087 }

\vfill\eject


\newsec{Introduction}

\def\FNphy{
     The $q$-KZ equation has been applied to the Thirring model in Ref. {\Rsm}
     and to the XXZ model or the vertex model in Ref. \refs{\Rdfjmn, \Riijmnt}.
}


 Recently,
the $q$-deformed Knizhnik-Zamolodchikov equation ($q$-KZ eq.),
more generally holonomic $q$-difference equation,
has been analyzed \refs{\Ra, \Rmi, \Rsm, \Rfr, \RmatI, \Rr}
and the remarkable relation between
the connection matrices of its solutions and the elliptic Boltzman weights
was investigated \refs{\Rfr, \Rakm}.

To study the $q$-KZ equation \foot{\FNphy},
an important and useful concept is the {\VOs},
which are $q$-analogue of the vertex operators in Ref. {\Rtk}.
First, the solutions of the $q$-KZ equation
can be constructed as the correlation functions of the {\VOs}
\refs{\Rsm, \Rfr}.
Moreover, we can understand naturally
the IRF type Yang-Baxter relation
satisfied by the connection matrices of $q$-KZ solutions
as the exchange relation for the {\VOs} \refs{\Rfr, \Rdjo}.
So far, explicit calculation of the exchange relation for the {\VOs}
has been done in Ref. {\Rfr} for the first nontrivial examples
and they conjectured that the connection matrices are
generally equivalent to the Boltzman weights in Ref. \refs{\Rabf, \Rdjmo}.
 The exchange relations of the {\VOs}
of spin $1/2$ or $k/2$ (of level $k$) were calculated
and applied to the vertex models in Ref. \refs{\Rdfjmn, \Riijmnt}.

 The aim of this paper is to generalize these results
to arbitrary spins.
We consider the $q$-KZ equation for the two point function
of the {\VOs} of spin $1/2$ and $j\in\ha\bZ_{\geq 0}$ and
we give the fundamental solutions and the connection matrices explicitly.


\newsec{Quantum affine algebra \Uqsl2 }

%
\noindent{\bf \S$\,$2.1.}~
 First we fix some notation.
The algebra {\Uqsl2} is generated by
$e_i$, $f_i$, invertible $k_i$ $(i =0,1)$ and $d$ with relations
\eqn\?{\eqalign{
k_i e_j k_i^{-1} &= q^{ a_{ij}} e_j ,\cr
k_i f_j k_i^{-1} &= q^{-a_{ij}} f_j ,\cr
e_i f_j -f_j e_i &= \d_{ij}{k_i - k_i^{-1} \over q-q^{-1}},\cr
}\qquad\eqalign{
\sum_{n=0}^{3}(-1)^n \Smat{ 3 \cr n \cr } e_i^{3-n} e_j e_i^n &=0 ,\cr
\sum_{n=0}^{3}(-1)^n \Smat{ 3 \cr n \cr } f_i^{3-n} f_j f_i^n &=0 ,\cr
}\qquad\eqalign{
[d,e_i]&= \d_{i,0} e_i,\cr
[d,k_i]&=  0          ,\cr
[d,f_i]&=-\d_{i,0} f_i,\cr
}}
and $k_0 k_1 = q^k$ with a level $ k \in \bC $,
where $q \in \bC$, $a_{11}=a_{00}=-a_{10}=-a_{01}=2$ and
\eqn\?{
\Smat{ n \cr m \cr } ={[n]! \over [n-m]! [m]! }, \qquad
[n]={q^n-q^{-n} \over q-q^{-1} }.
}

%
The algebra {\Uqsl2} is a Hopf algebra
with the comultiplication $\D$, the antipode $S$ and the co-unit $\e$
\eqn\?{\eqalign{
\D(e_i) &= e_i \otimes k_i + 1        \otimes e_i ,\cr
\D(k_i) &= k_i \otimes k_i                        ,\cr
\D(f_i) &= f_i \otimes 1   + k_i^{-1} \otimes f_i ,\cr
\D(d  ) &= d   \otimes 1   + 1        \otimes d   ,\cr
}\qquad\eqalign{
S(e_i)  &=-e_i k_i^{-1} ,\cr
S(k_i)  &= k_i^{-1}     ,\cr
S(f_i)  &=-k_i f_i      ,\cr
S(d  )  &=-d            ,\cr
}\qquad\eqalign{
\e(e_i) &= 0 ,\cr
\e(k_i) &= 1 ,\cr
\e(f_i) &= 0 ,\cr
\e(d  ) &= 0 .\cr
}}

\vskip 2mm
\noindent{\bf \S$\,$2.2.}~
 Let $V_j$ be the Verma module over {\Uqsl2},
generated by the highest weight vector $\|j\>$, such that
$e_i\|j\> = 0 $,
$k_1\|j\> = q^{2j}\|j\>$ and
$d  \|j\> =-h_j   \|j\>$ with $h_j=j(j+1)/\k$, $\k=k+2$.
 The dual module $V_j^*$ is generated by $ \<j\|$ which satisfies
$\<j\|e_i = 0 $,
$\<j\|k_1 = q^{2j}\<j\|$ and
$\<j\|d   =-h_j   \<j\|$.
The bilinear form
$V_j^* \otimes V_j \-> \bC  $ is uniquely defined by
$\<j\|j\> = 1 $ and
$\big(\<u\|a\big)\|v\> = \<u\|\big(a\|v\>\big) $
for any $\<u\|\in V_j^*$,  $\|v\>\in V_j$ and $a\in${\Uqsl2}.

 A null vector $\|\chi\> \in V_j$
(of grade $N$ and charge $Q$ ) is defined by
$e_i\|\chi\>=0$,
$k_1\|\chi\>=q^{2(j+Q)}\|\chi\>$ and
$d  \|\chi\>=-({h_j+N})\|\chi\>$.
 A null vector $\<\chi\|\in V_j^*$ is defined in a similar manner.


For $2j \in \bZ _{\ge 0}$,
we have a $2j+1$ dimensional centerless irreducible representation of {\Uqsl2},
$V_j(z)= \oplus^{2j}_{m=0} \bC (q,z)\|j,m\>$, which is defined by
\eqn\Efinite{\eqalign{
e_1\|j,m\> &=   [2j-m+1]\,\|j,m-1\> ,\cr
k_1\|j,m\> &=q^{2(j-m)}   \|j,m  \> ,\cr
f_1\|j,m\> &=   [   m+1]\,\|j,m+1\> ,\cr
}\qquad\eqalign{
e_0\|j,m\> &=z     [   m+1]\,\|j,m+1\> ,\cr
k_0\|j,m\> &=  q^{-2(j-m)}   \|j,m  \> ,\cr
f_0\|j,m\> &=z^{-1}[2j-m+1]\,\|j,m-1\> ,\cr
}}
and $d$ acts as $d =-h_j + z {d \over dz}$.


\newsec{ The $q$-vertex operators and their two point functions }

%
\def\FNtype-II{\rm
  Our  {\VOs} correspond to the type-II vertex operators
  in Ref. \refs{\Rdjo, \Rdfjmn, \Riijmnt, }.
  The relation between our {\VO}
  $\Phi_{j_2,}{}_{m_2}^{j_1+j_2-j_3}(z): V_{j_1} \-> V_{j_3}$
  and their one
  $\Phi^{j_3}_{j_2,j_1}(z) : V_{j_2}(z) \otimes V_{j_1} \-> V_{j_3}$ is
  $$
    \Phi_{j_2,}{}_{m_2}^{j_1+j_2-j_3}(z)
  = \Phi^{j_3}_{j_2,j_1}(z) \Big( \,\,\|j_2,m_2\>,\,\,  * \,\,\Big).
  $$
  The $\Phi^{j_3}_{j_2,j_1}(z)$ has the intertwining property,
  $a \,\Phi^{j_3}_{j_2,j_1}(z) = \Phi^{j_3}_{j_2,j_1}(z) \,\D(a)$,
  for all $a \in ${\Uqsl2}.
  Note that the coproduct is also slightly different.
}
\def\FNindex{\rm
  The upper index $\a$, which specify the modules
  on which the {\VO} acts, will be sometimes suppressed.
}
\def\FNsign{\rm
  The sign difference of $d$ in \S 2.2 and \S 3.1
  comes from the difference in meaning of $z$.
  In \S 2.2, $z$ is a generator such that $[d,z]=1$,
  on the other hand  $z$ is just a variable i.e. $[d,z]=0$, in \S 3.1.
}
%
\noindent{\bf \S$\,$3.1.}~
The {\VO}, $\Phi_{j_2}(z):V_{j_1}\->V_{j_3}$ , of spin $j_2\in \bC$
is defined by the transformation property under the adjoint action
\refs{\Rfr, \Rsm, \Rfl},
or equivalently as a intertwiner (see Appendix A).

\noindent{\underbar{\bf Definition}}.~{\it
 For $2j_2\in\bZ_{\geq 0}$, the {\VO} {\rm\foot{\FNtype-II}},
$\Phi_{j_2,}{}_{m_2}^{\a_2}(z):V_{j_1}\-> V_{j_3}$
with $j_3=j_1+j_2-\a_2$ and $0 \leq m_2 \leq 2j_2$,
is defined explicitly as follows {\rm\foot{\FNindex}}
}
\eqn\Eqvo{\eqalign{
e_1\Phi_{j,m}(z)&=[2j-m+1]\,\Phi_{j,m-1}(z)\,k_1+      \Phi_{j,m}(z)\,e_1,\cr
k_1\Phi_{j,m}(z)&=q^{2(j-m)}\Phi_{j,m  }(z)\,k_1                         ,\cr
f_1\Phi_{j,m}(z)&=[   m+1]\,\Phi_{j,m+1}(z)+q^{-2(j-m)}\Phi_{j,m}(z)\,f_1,\cr
\apar
e_0\Phi_{j,m}(z)&=z     [   m+1]\,\Phi_{j,m+1}(z)\,k_0+\Phi_{j,m}(z)\,e_0,\cr
k_0\Phi_{j,m}(z)&=  q^{-2(j-m)}   \Phi_{j,m  }(z)\,k_0                   ,\cr
f_0\Phi_{j,m}(z)&=z^{-1}[2j-m+1]\,\Phi_{j,m-1}(z)
                                            +q^{2(j-m)}\Phi_{j,m}(z)\,f_0,\cr
}}
{\it and} \foot{\FNsign}
\eqn\?{
d\,\Phi_{j,m}(z)=-z{d\over dz}\,\Phi_{j,m}(z)+\Phi_{j,m}(z)\,d.
}

  The existence conditions for the $q$-vertex operator
$\Phi_{j,}{}_{m}(z)$ with a general spin $j\in\bC $
were analyzed in Ref. {\RayII}, and we will review them in Appendix A.
In the case that $q$ is not a root of unity,
the existence conditions are essentially the same as those of $q=1$ {\RayI}.
For integrable representations of general quantum affine algebras,
the complete results on existence and uniqueness were given in Ref. {\Rdjo}.


 From the $ k_1 $ and $ d $ commutation relations,
the ground state matrix element of the {\VO} can be determined
up to normalization, and we normalize it as
\eqn\?{
\<j_3\|\Phi_{j_2,}{}_{m_2}^{\a_2}(z)\|j_1\> =
\d^{j_3-j_1}_{j_2-m_2} \d^{j_3-j_1}_{j_2-\a_2} z^{ h_3-h_1 },
}
where $h_n=h_{j_n}$.
The other matrix elements for the descendant fields
can be uniquely determined.


We can in principle derive the arbitrary $N$ point functions
by using this one point function
and the $q$-operator product expansion ($q$-OPE).
In Appendix B, we will present the $q$-OPE of a spin $1/2$ {\VO}
and the two point functions
$\<j_4\|\,\Phi_{\ha,m_3}(z_3)\,\Phi_{j_2,m_2}(z_2)\,\|j_1\> $ and
$\<j_4\|\,\Phi_{j_2,m_2}(z_2)\,\Phi_{\ha,m_3}(z_3)\,\|j_1\> $.

\vskip 2mm
\noindent{\bf \S$\,$3.2.}~
As we will discuss in Appendix B,
the above two point functions have a complicated form,
for example each coefficient of $z^n$ can not be factorised.
But they can be simplified by dividing by a function
\eqn\EgOPE{
g(x)=(1-q^{k+2}{ [2j] \over [2][k+2] }x + \iO (x^2) ),
\qquad
x=z_2/z_3.
}
Here we list the leading terms
\eqn\Etwo{\eqalign{
&\opeOO =z_3^{h_4-h_4^-}z_2^{h_4^- -h_1}\cr
&\,\hskip 3cm \times
   \{1+ p^{-m_1-m_2} x{[2m+m_1+m_2+1]_p[2m_1]_p\over [2m+m_1-m_2+1]_p}
                                                 +\iO (x^2) \}g(x),\cr
\apar
&\opeOI =-z_3^{h_4-h_4^-}z_2^{h_4^- -h_1} x
                          p^{-m_2}{ [2m_1]_p\over[2m+m_1-m_2+1]_p}\cr
&\,\hskip 3cm \times
   \{1+ p^{-m_1-m_2} x{[2m+m_1+m_2+1]_p[2m_1+1]_p\over [2m+m_1-m_2+2]_p}
                                                +\iO (x^2) \}g(x),\cr
}}
\eqn\?{\eqalign{
&\opeIO =-z_3^{h_4-h_4^+}z_2^{h_4^+ -h_1}
                         p^{-m_1}{ [2m_2]_p\over[-2m-m_1+m_2]_p }\cr
&\,\hskip 3cm \times
   \{1+ p^{-m_1-m_2} x{[-2m+m_1+m_2]_p[2m_2+1]_p\over [-2m-m_1+m_2+1]_p}
                                                +\iO (x^2) \}g(x),\cr
\apar
&\opeII =z_3^{h_4-h_4^+}z_2^{h_4^+ -h_1}\cr
&\,\hskip3cm \times
   \{1+ p^{-m_1-m_2}x{[-2m+m_1+m_2]_p[2m_2]_p\over [-2m-m_1+m_2]_p}
                                               +\iO (x^2) \}g(x),\cr
}}
and
\eqn\?{\eqalign{
&\opeOOr =z_2^{h_4-h_1^+}z_3^{h_1^+ -h_1}\cr
&\,\hskip 2cm \times\{1+ p^{-m_1-m_2} x^{-1}
{ [-2m+m_1+m_2]_p[2m_1]_p\over [-2m+m_1-m_2]_p}
+\iO (x^{-2}) \}g(x^{-1}),\cr
\apar
&\opeOIr =-z_2^{h_4-h_1^+}z_3^{h_1^+ -h_1}
p^{-m_2}{ [2m_1]_p\over[-2m+m_1-m_2]_p }\cr
&\,\hskip 2cm \times\{1+ p^{-m_1-m_2} x^{-1}
{ [-2m+m_1+m_2]_p[2m_1+1]_p\over [-2m+m_1-m_2+1]_p}
+\iO (x^{-2}) \}g(x^{-1}),\cr
}}
\eqn\?{\eqalign{
&\opeIOr =-z_2^{h_4-h_1^-}z_3^{h_1^- -h_1} x^{-1}
p^{-m_1}{ [2m_2]_p\over[2m-m_1+m_2+1]_p}\cr
&\,\hskip 2cm \times\{1+ p^{-m_1-m_2} x^{-1}
{ [2m+m_1+m_2+1]_p[2m_2+1]_p\over [2m-m_1+m_2+2]_p}
+\iO (x^{-2}) \}g(x^{-1}),\cr
\apar
&\opeIIr =z_2^{h_4-h_1^-}z_3^{h_1^- -h_1}\cr
&\,\hskip 2cm \times\{1+ p^{-m_1-m_2} x^{-1}
{ [2m+m_1+m_2+1]_p[2m_2]_p\over [2m-m_1+m_2+1]_p}
+\iO (x^{-2}) \}g(x^{-1}),\cr
}}
here and below we use the following notations,
$[n]_p={(p^n-1)/(p-1)}$, $p=q^{-2\k}$,
$m  =-(j_1    +j_4+1   )/ 2\k $,
$m_1=( j_1+j_2-j_4+\ha )/ 2\k = M/ 2\k $,
$m_2=(-j_1+j_2+j_4+\ha )/ 2\k =\M/ 2\k $ and
$h_n^{\pm}=h_{j_n\pm\ha}$.

More complete expressions for the  two point functions
will be given in \S 5.1 by solving the $q$-KZ equation.


\newsec{The solutions and their connection formula for the $q$-KZ equation  }

%
\noindent{\bf \S$\,$4.1.}~
Arbitrary $N$ point functions of {\VOs} satisfy the $q$-KZ equation
\refs{\Rfr}.
For the two point function of the {\VOs} of spin $1/2$ and $j$,
$\<j_4\|\,\Phi_{\ha,M_3}(z_3)\,\Phi_{j_2,M_2}(z_2)\,\|j_1\> $,
the $q$-KZ equation is written as a $2 \times 2$ block diagonal
$R$ matrix $R(z_2/z_3):
V_{j_2}(z_2)\otimes V_{\ha}(z_3) \-> V_{j_2}(z_2)\otimes V_{\ha}(z_3)$,
defined by $R(z_2/z_3)\D'_{z_2,z_3}=\D_{z_2,z_3}R(z_2/z_3)$ (Appendix C).

Let $M=M_2+M_3$ , $M+\M=2j_2+1$ and $x=z_2/z_3$.

Up to normalization, this $q$-KZ equation is
\eqn\Eqkz{\eqalign{
\Pmat{ p^{-m}\tPsi_0(px) \cr
       p^{ m}\tPsi_1(px) \cr }
&=\tR(x)
\Pmat{ \tPsi_0(x) \cr
       \tPsi_1(x) \cr },\cr
\apar
\tR(x)=
\Pmat{ \tR_0^0(x) & \tR_0^1(x) \cr
       \tR_1^0(x) & \tR_1^1(x) \cr }
&={ 1 \over  1 - x p^{m_1+m_2} }
\Pmat{         p^{ m_1}-xp^{m_2}  & p^{m_2}(p^{-m_2}- p^{m_2}) \cr
      xp^{m_1}(p^{-m_1}- p^{m_1}) &         p^{ m_2}-xp^{m_1}  \cr
},
}}
where $p$, $m$ and $m_i$ are the same as in \S 3.2.


The equation {\Eqkz} can be solved easily,
and the solutions are given by the $q$-hypergeometric function
\eqn\?{
F_p(a,b,c;x)=\sum_{n=0}^{\infty}{(a)_n (b)_n \over (c)_n (1)_n } x^n,
}
where $(a)_n=[a]_p[a+1]_p \cdots [a+n-1]_p$.

Two fundamental solutions
$\tPsi_{(+)}{}^0_i(x)$ and $\tPsi_{(+)}{}^1_i(x)$
in the region $x \ll 1$ are
\eqn\?{\eqalign{
\tPsi_{(+)}{}^0_0(x)&=
-F(-2m+m_1+m_2,2m_2+1,-2m-m_1+m_2+1;p^{-m_1-m_2}x)\cr
&\, \hskip5.5cm \times
p^{-m_1} {[2m_2]_p \over [-2m-m_1+m_2]_p} x^{m_2-m} ,\cr
\tPsi_{(+)}{}^0_1(x)&=
 F(-2m+m_1+m_2,2m_2  ,-2m-m_1+m_2  ;p^{-m_1-m_2}x) x^{m_2-m} ,\cr
\tPsi_{(+)}{}^1_0(x)&=
 F(2m+m_1+m_2+1,2m_1  ,2m+m_1-m_2+1  ;p^{-m_1-m_2}x) x^{m+m_1} ,\cr
\tPsi_{(+)}{}^1_1(x)&=
-F(2m+m_1+m_2+1,2m_1+1,2m+m_1-m_2+2;p^{-m_1-m_2}x)\cr
&\, \hskip5cm \times
p^{-m_2} {[2m_1]_p \over [2m+m_1-m_2+1]_p} x^{m+m_1+1}.\cr
}}
And the other two fundamental solutions
$\tPsi_{(-)}{}^0_i(x)$ and $\tPsi_{(-)}{}^1_i(x)$
in the region $x \gg 1$ are
\eqn\?{\eqalign{
\tPsi_{(-)}{}^0_0(x)&=
-F(2m+m_1+m_2+1,2m_2+1,2m-m_1+m_2+2;p^{-m_1-m_2+1}x^{-1})\cr
&\, \hskip4.5cm \times
p^{2m-m_1+1} {[2m_2]_p \over [2m-m_1+m_2+1]_p} x^{-m-m_2-1},\cr
\tPsi_{(-)}{}^0_1(x)&=
 F(2m+m_1+m_2+1,2m_2 ,2m-m_1+m_2+1 ;p^{-m_1-m_2+1}x^{-1})x^{-m-m_2},\cr
\tPsi_{(-)}{}^1_0(x)&=
 F(-2m+m_1+m_2,2m_1  ,-2m+m_1-m_2  ;p^{-m_1-m_2+1}x^{-1})x^{m-m_1} ,\cr
\tPsi_{(-)}{}^1_1(x)&=
-F(-2m+m_1+m_2,2m_1+1,-2m+m_1-m_2+1;p^{-m_1-m_2+1}x^{-1})\cr
&\, \hskip5.5cm \times
p^{-2m-m_2} {[2m_1]_p \over [-2m+m_1-m_2]_p} x^{m-m_1} .\cr
}}


\vskip 2mm
\noindent{\bf \S$\,$4.2.}~
The complete $q$-KZ equation with the correct normalization is
\eqn\EqKZ{
\Pmat{ p^{-m}\Psi_{0}(px) \cr
       p^{ m}\Psi_{1}(px) \cr }
= R(x)
\Pmat{ \Psi_{0}(x) \cr
       \Psi_{1}(x) \cr },
}
where $R(x)$ is the image of the universal $R$-matrix
which satisfies the crossing relation (see Appendix C).
This $R(x)$ is given by
\eqn\Ef{
R(x) = f(x) \tR(x), \qquad
f(x) =\prod_{n \geq 0}
             {(1-xq^{2j_2+3+4n})(1-xq^{-2j_2+1+4n})\over
              (1-xq^{2j_2+1+4n})(1-xq^{-2j_2+3+4n})     },
}
in the region $x \ll 1$.


The relation of the $\Psi(x)$'s and the previous $\tPsi(x)$'s is
\eqn\Efg{
\Psi_{i}(x)=g(x)\tPsi_{i}(x), \qquad
g(xp)=f(x)g(x).
}
The factor $g(x)$ in the regions $x \ll 1$ and $x \gg 1$ are
\eqn\Eg{\eqalign{
g_{+}(x)&=\prod_{n,m \geq 0}
  {(1-xq^{2j_2+1+4n}p^m)(1-xq^{-2j_2+3+4n}p^m)\over
   (1-xq^{2j_2+3+4n}p^m)(1-xq^{-2j_2+1+4n}p^m)     },
\cr
g_{-}(x)&=\prod_{n,m \geq 0}
  {(1-x^{-1}q^{-2j_2-3-4n}p^{m+1})(1-x^{-1}q^{2j_2-1-4n}p^{m+1})\over
   (1-x^{-1}q^{-2j_2-1-4n}p^{m+1})(1-x^{-1}q^{2j_2-3-4n}p^{m+1})     },
}}
respectively.

%
\def\FNhyper{For the proof of this formula, see Ref. {\Rmi} for example.}
%
%
\vskip 2mm
\noindent{\bf \S$\,$4.3.}~
The connection formula for the $q$-hypergeometric function
$F_p(a,b,c;x)$ is \foot{\FNhyper}
\eqn\?{\eqalign{
F_p(a,b,c;x)
&={\G_p(c)\G_p(b-a)\T(p^ax,p)\over \G_p(b)\G_p(c-a)\T(x,p)}
F_p(a,a-c+1,a-b+1,p^{c+1-a-b}x^{-1}) \cr
&+{\G_p(c)\G_p(a-b)\T(p^bx,p)\over \G_p(a)\G_p(c-b)\T(x,p)}
F_p(b,b-c+1,a-b+1,p^{c+1-a-b}x^{-1}),
}}
where
\eqn\Egt{\eqalign{
\G_p(a)&=(1-p)^{1-a}\prod_{n\geq 0}{(1-p^{n+1})\over (1-p^{n+a})},
\cr
\T(x,p)&=\prod_{n\geq 0}(1-p^{n+1})(1-xp^n)(1-x^{-1}p^{n+1}).
}}
Using this formula, we have

\noindent\underbar{\bf Proposition I}.~{\it
The connection formulas for the solutions of the $q$-KZ equation
{\rm{\Eqkz}} and for the normalization factor $g(x)$ are
\eqn\EconI{
\tPsi_{(+)}{}^{\a}_i(x) =\sum_{\b }
\tPsi_{(-)}{}^{\b }_i(x)  C^{\a}_{\b }(x),
}
\eqn\?{\eqalign{
C^0_0(x)&=
{\G_p(2m+m_1-m_2+1)\G_p(2m-m_1+m_2+1) \over
 \G_p(2m-m_1-m_2+1)\G_p(2m+m_1+m_2+1)}
{\T(p^{m_1-m_2}x)\over \T(p^{-m_1-m_2}x) } x^{2m_1},\cr
C^0_1(x)&=-
{\G_p(-2m+m_1-m_2  )\G_p(2m+m_1-m_2+1) \over
 \G_p(      -2m_2+1)\G_p(  2m_1      )}
{\T(p^{2m+1}x)\over \T(p^{-m_1-m_2}x) } p^{-m_2} x^{2m+m_1+m_2+1},\cr
C^1_0(x)&=-
{\G_p(-2m-m_1+m_2)\G_p(2m-m_1+m_2+1) \over
 \G_p(  -2m_1+1  )\G_p(      2m_2  )}
{\T(p^{-2m}x)\over\T(p^{-m_1-m_2}x) } p^{-m_1} x^{-2m+m_1+m_2},\cr
C^1_1(x)&=
{\G_p(-2m+m_1-m_2)\G_p(-2m-m_1+m_2) \over
 \G_p(-2m-m_1-m_2)\G_p(-2m+m_1+m_2)}
{\T(p^{m_2-m_1}x)\over \T(p^{-m_1-m_2}x) } x^{2m_2},\cr
}}
and
\eqn\EconII{
g_{+}(x)=g_{-}(x) C_g(x),\qquad
C_g(x)=\prod_{n \geq 0}
{\T(xq^{2j_2+1+4n}) \T(xq^{-2j_2+3+4n}) \over
 \T(xq^{2j_2+3+4n}) \T(xq^{-2j_2+1+4n})       }.
}
} 
Note that $C^{\a}_{\b}(x)$ and $C_g(x)$ are the pseudo-constant,
e.g. $C_g(px)=C_g(x)$.


\newsec{Exchange relation for the \VOs }

%
\noindent{\bf \S$\,$5.1.}~
Comparing the two point function in \S 3.2 and $\tPhi(x)$ in \S 4.1,
we have

\noindent{\underbar{\bf Proposition II}}.~{\it 
The relations between
the solutions of the reduced $q$-KZ equation in {\rm\S 4.1}
with the two point functions in {\rm\S 3.2} are
}
\eqn\?{\eqalign{
&\Pmat{\opeOO & \opeIO \cr
       \opeOI & \opeII \cr }\cr
&\, \hskip6cm
=x^{-j_2/2\k} z_2^{h_4-h_1}g_{+}(x)
\Pmat{ \tPsi_{(+)}{}^0_0(x) &\tPsi_{(+)}{}^1_0(x) \cr
       \tPsi_{(+)}{}^0_1(x) &\tPsi_{(+)}{}^1_1(x) \cr },\cr
&\Pmat{\opeOOr & \opeIOr \cr
       \opeOIr & \opeIIr \cr}\cr
&\, \hskip3cm
=x^{j_2/2\k} z_2^{h_4-h_1}g_{-}(px)
\Pmat{p^{-m+m_1}\tPsi_{(-)}{}^0_0(px)&p^{-m+m_2}\tPsi_{(-)}{}^1_0(px)\cr
      p^{ m+m_1}\tPsi_{(-)}{}^0_1(px)&p^{ m+m_2}\tPsi_{(-)}{}^1_1(px)\cr}.
}}
\noindent {\it Proof.}~
Both sides of the above equations satisfy the same $q$-KZ equations.
So we need only to compare the leading terms.
{}From $g_+(x) =g(x)     +{\it O}(x^2   )$
and  $g_-(px)=g(x^{-1})+{\it O}(x^{-2})$,
we obtain the required relation.
\hfill Q.E.D.


{}From the difference equation for $g(x)$ and $\tPhi(x)$
and from the Proposition-I,-II, we have

\noindent\underbar{\bf Theorem}.~ {\it 
The exchange relation for the {\VOs} of
spin $1/2$ and arbitrary spin $j_2 \in \ha\bZ_{\geq 0}$ is as follows
\eqn\Etheorem{
\cR_{ij}^{kl}({z\over w})
  \Phi_{\ha,}{}_{k}^{\a}(w) \Phi_{j_2,}{}_{l}^{\b}(z)
= \Phi_{j_2,}{}_{i}^{\g}(z) \Phi_{\ha,}{}_{j}^{\d}(w)
\cC^{\a \b}_{\g\d}({z\over w},\la),
}
with
\eqn\Er{
\Pmat{\cR_{M  ,}^0{}_0^M (x) &\cR_{M  ,}^1{}_0^{M-1} (x) \cr
      \cR_{M-1,}^0{}_1^M (x) &\cR_{M-1,}^1{}_1^{M-1} (x) \cr }
={f(x)\over x-q^{M+\M}}
\Pmat{       (xq^{ M}-q^{\M}) &q^M( q^{-\M}-q^{\M})   \cr
      xq^{\M}( q^{-M}-q^{ M}) &    xq^{ \M}-q^{ M}    \cr },
}
\eqn\Ec{
\Pmat{\cC_{M  ,}^0{}_0^M(x,\la)&\cC_{M  ,}^1{}_0^{M-1}(x,\la)\cr
      \cC_{M-1,}^0{}_1^M(x,\la)&\cC_{M-1,}^1{}_1^{M-1}(x,\la)\cr}
= x^{-{j_2/ \k}}  C_g(x)
\Pmat{p^{-m_1} C^0_0 (x) & p^{-m_1} C^1_0 (x) \cr
      p^{-m_2} C^0_1 (x) & p^{-m_2} C^1_1 (x) \cr } ,
}
where $M=\a+\b=\g+\d=i+j=k+l$, $\la =2j_4+1$ and
$f(x)$, $C^{\a}_{\b}(x)$ and $C_g(x)$ are
as given in {\rm {\Ef}}, {\rm{\EconI,}} and {\rm{\EconII}} respectively.
}

\noindent{\it Proof.}~
{}From the intertwining property of the $R$-matrix,
it is obvious that the Theorem holds
not only for
the lowest  state $\<j_4\|\in V^*_{j_4}$ and
the highest state $\|j_1\> \in V_{j_1}$
but also for arbitrary states in $V^*_{j_4}$ and $V_{j_1}$.
\hfill Q.E.D.


\vskip 2mm
\noindent{\bf \S$\,$5.2.}~
If we denote the {\VO} and the connection matrix as
\eqn\?{\eqalign{
\Phi\Smat{j_2 \cr j_3 \,\,\, j_1 \cr}_{m_2}(z)
&=\Phi_{j_2,}{}_{m_2}^{j_1+j_2-j_3}(z)\quad:\quad V_{j_1}\->V_{j_3} ,\cr
C_{\ha j_2}\Smat{j_4 \,\,\, j' \cr j \,\,\, j_1 \cr}
({z\over w})
&=\cC^{j+\ha-j_4\, ,}_{j'+j_2-j_4 ,}{}^{j_1+j_2-j}_{j_1+\ha-j'}
({z\over w},\la),
}}
then the exchange relation {\Etheorem} can be written as
\eqn\?{
\cR_{ij}^{kl}({z\over w})
  \Phi\Smat{\ha \cr j_4 \,\,\,\, j \,\cr}_{k}(w)
  \Phi\Smat{j_2 \cr j     \,\,\, j_1 \cr}_{l}(z)
= \Phi\Smat{j_2 \cr j_4   \,\,\, j'  \cr}_{i}(z)
  \Phi\Smat{\ha \cr j'  \,\,\,\, j_1 \cr}_{j}(w)
C_{\ha j_2}\Smat{j_4 \,\,\, j' \cr j \,\,\, j_1 \cr}
({z\over w}).
}
{}From the fact that the $R$-matrix $\cR_{ij}^{kl}({z\over w}) $
satisfies the Yang-Baxter equation, the connection matrix
$C_{\ha j_2}\Smat{j_4 \,\,\, j' \cr j \,\,\, j_1 \cr}
({z\over w})$
obeys the IRF type Yang-Baxter relation \refs{\Rfr, \Rdjo}.
Moreover if we denote
$[n]_{\vt}=p^{-n/2\k}\T(p^{n/\k})$,
$x=q^{2j_2+1-2u}$, then
\eqn\?{\eqalign{
&\Pmat{\cC_{M  ,}^0{}_0^M(x,\la) &\cC_{M  ,}^1{}_0^{M-1}(x,\la)\cr
       \cC_{M-1,}^0{}_1^M(x,\la) &\cC_{M-1,}^1{}_1^{M-1}(x,\la)\cr }
={ C_g(x) x^{-j_2/ \k} \over [\la+M-\M]_{\vt}[u-M-\M]_{\vt} } \cr
\apar
&\, \hskip 2cm \times
\Pmat{ \tilde\a_1  &\,          \cr
       \,          &\tilde\a_2  \cr }
\Pmat{[\la-\M]_{\vt}[u-\M    ]_{\vt}&[   - M]_{\vt}[u-\M+\la]_{\vt}\cr
      [    \M]_{\vt}[u- M-\la]_{\vt}&[\la+ M]_{\vt}[u- M    ]_{\vt}\cr}
\Pmat{ \a_1  & \,     \cr
       \,    & \a_2   \cr }^{-1},  \cr}
}
where
\eqn\?{
C_g(x)
=\prod_{n \geq 0}{[u-M-\M  -2n]_{\vt}[u-2-2n]_{\vt}
             \over[u-M-\M-1-2n]_{\vt}[u-1-2n]_{\vt}},
}
\eqn\?{\eqalign{
      \a_1 &= {\G_p(2m_1) \G_p( 2m-m_1-m_2+1)\over \G_p( 2m+m_1-m_2+1)}
                                                             x^{-m-m_1},\cr
      \a_2 &= {\G_p(2m_2) \G_p(-2m-m_1-m_2  )\over \G_p(-2m-m_1+m_2  )}
                                                     x^{ m-m_2}p^{-m_2},\cr
\tilde\a_1 &= {\G_p(2m_1) \G_p(-2m-m_1-m_2  )\over \G_p(-2m+m_1-m_2  )}
               x^{-m+m_1}p^{-m_1},\cr
\tilde\a_2 &= {\G_p(2m_2) \G_p( 2m-m_1-m_2+1)\over \G_p( 2m-m_1+m_2+1)}
               x^{ m+m_2},\cr
}}
this connection matrix is equivalent to
the elliptic Boltzman weight of IRF type
obtained by the fusion procedure in Ref. {\Rdjmo}.


\newsec{Conclusion}


 We have given explicitly the two point functions
and the exchange relations for
the {\VOs} whose spins are $1/2$ and arbitrary $j\in\ha\bZ_{\geq 0}$.
We also confirmed that the connection matrix is equivalent to
the elliptic Boltzman weight of IRF type
obtained by the fusion procedure in Ref. {\Rdjmo}.

 We expect that the exchange relation for the two arbitrary spin {\VOs}
will be given by an analogous fusion procedure.
Our method essentially relies on the connection formula of
$q$-hypergeometric function
and it is applicable only to the case
when the number of intermediate channels is at most two.
A more promising approach to the connection problem is
to use the integral formula.

  Recently, a free field realization for {\Uqsl2} was constructed
\refs{\RmatII, \Rabg, \Rsh}.
The free field realization will give a powerful tool {\Raty} to calculate
the integral formulas \refs{\Ra, \Rmi, \RmatI, \Rr}
for the $q$-KZ solution and to solve their connection problem.


\bigbreak\centerline{\bf Acknowledgment.}

 This work has been carried out in collaboration with Y. Yamada.
I am grateful to him for help in many ways.
 I would like to thank M. Jimbo, J. Shiraishi
and the members of KEK theory group
for valuable discussions.
I would also like to thank to N. A. McDougall
for a careful reading of the manuscript.


\appendix{A}{The existence condition for the $q$-vertex operator
of arbitrary spin }

%
\noindent{\bf \S$\,$A.1.}~
Here we briefly review the result of our previous paper \refs{\RayII}.
For an arbitrary spin $j\in\bC$,
the {\VO} $\Phi_{j_2}(z):V_{j_1}\-> V_{j_3}$ is defined by the
transformation property under the adjoint action of {\Uqsl2}
\refs{\Rfr, \Rsm, \Rfl}, such that
${\rm ad} (a)\,\Phi(z)\equiv\sum_k a_k^{1}\,\Phi(z)\,S(a_k^{2})$
with $\Delta(a)=\sum_k a_k^{1} \otimes a_k^{2}$.

\noindent{\underbar{\bf Definition}}.~{\it
The {\VO} $\Phi_{j_2}(z):V_{j_1}\-> V_{j_3}$ is defined as
\eqn\?{
{\rm ad} (a) \,\Phi(z) =\r(a) \,\Phi(z),
}
for ${}^{\forall} a \in${\Uqsl2}, where
$\r$ is a certain representation of {\Uqsl2}.
}

\noindent
The following intertwining property is equivalent to above definition;
\eqn\?{
a\,\Phi(z) = \sum_k \r(a_k^{1})\,\Phi(z)\,a_k^{2}.
}

 For $\rho$, if we take the contravariant difference representation
defined by
\eqn\IIrep{\eqalign{
\rho(e_1) &=          x  [2j - x {d \over dx}] ,\cr
\rho(k_1) &= q^{  2(j - x {d \over dx})}       ,\cr
\rho(f_1) &= {1 \over x} [     x {d \over dx}] ,\cr
}\qquad\eqalign{
\rho(e_0) &= {z \over x} [     x {d \over dx}] ,\cr
\rho(k_0) &= q^{ -2(j - x {d \over dx})}       ,\cr
\rho(f_0) &= {x \over z} [2j - x {d \over dx}] ,\cr
}\qquad\eqalign{
\,\cr
\rho(d  ) &=   {      - z {d \over dz} }, \cr
\,
}}
then the {\VO} $ \Phi_{j}(z,x) $
of spin $j$ can be defined explicitly as follows \refs{\RayII}
\eqn\?{\eqalign{
e_i\Phi_j(z,x)&=\r(e_i)\,\Phi_j(z,x)\,k_i + \Phi_j(z,x)\,e_i ,\cr
k_i\Phi_j(z,x)&=\r(k_i)\,\Phi_j(z,x)\,k_i ,\cr
f_i\Phi_j(z,x)&=\r(f_i)\,\Phi_j(z,x)
               +\r(k_i^{-1})\,\Phi_j(z,x)\,f_i,\cr
}}
and
\eqn\?{
d\,\Phi_j(z,x)=\r(d)\,\Phi_j(z,x) +\Phi_j(z,x)\,d.
}
Here $[n]$ denotes the $q$ integer $[n]=( q^n-q^{-n} )/( q-q^{-1} )$,
so the $\rho (J)$'s are the difference operators,
e.g. $q^{x{d\over dx}}f(x)=f(qx)$ for any function $f(x)$.

And we normalize the ground state matrix element of the {\VO} as
\eqn\?{
\<j_3\|\Phi_{j_2}(z,x)\|j_1\>=z^{h_3-h_1} x^{j_1+j_2-j_3}.
}
 For $2j+1\in\bZ_{>0}$,
if we set $\Phi_j(z,x)=\sum_{m=0}^{2j} \Phi_{j,m}(z) x^m$,
then we have the previous definition for the {\VOs} $\Phi_{j,m}(z)$ in \S3.1.

\vskip 2mm
\noindent{\bf \S$\,$A.2.}~
The {\VO} $\Phi_{j_2}(z,x):V_{j_1}\-> V_{j_3}$ exists if and only if
\eqn\?{
\<j_3\|\Phi_{j_2}(z,x)\|\chi_1\>=\<\chi_3\|\Phi_{j_2}(z,x)\|j_1\> = 0,
}
for all the null vectors $\|\chi_1\> \in V_{j_1}$ and
                         $\<\chi_3\|\in V_{j_3}^*$.

For $r,s\in\bZ$ and $j_n\in\bC$, let
\eqn\?{\eqalign{
&f_{r,s}(j_1,j_2,j_3)=
\prod_{n=0}^{r-1} \prod_{m=0}^{s} [ j_1 + j_2 - j_3 - n + m\k ]
\prod_{n=1}^{r  } \prod_{m=1}^{s} [-j_1 + j_2 + j_3 + n - m\k ],
\cr
&f_{r,s}(j_1,j_2,j_3)=
\prod_{n=0}^{-r-1} \prod_{m=0}^{-s-1} [-j_1 + j_2 + j_3 - n + m\k ]
\prod_{n=1}^{-r  } \prod_{m=1}^{-s-1} [ j_1 + j_2 - j_3 + n - m\k ].
}}
for the case (i) $r>0$ and $s \geq 0$ or
(ii) $r<0$ and $s<0$ , respectively.
Then we have

\noindent{\underbar{\bf Theorem}}.\refs{\RayII}~{\it
The existence conditions for the {\VOs} $\Phi_{j_2}(z,x):V_{j_1}\-> V_{j_3}$
are given as follows.\hfill\break
\noindent {\rm (I)}. For the rational level $\k=p/q$,
with the coprime integers $p$ and $q$,
and $2j_n+1=r_n-s_n\k$ with $0<r_n<p$ and $0\leq s_n <q$, $(n=1,2,3)$,
the {\VO} exists if and only if
\eqn\?{
f_{r_1,s_1}(j_1,j_2,j_3)=f_{r_1-p,s_1-q}(j_1,j_2,j_3)=0,
}
\noindent {\rm (II)}. For the generic level,
the {\VO} exists if and only if
\eqn\?{
 \sum_{r_1,s_1\in\bZ}f_{r_1,s_1}(j_1,j_2,j_3)\d^{2j_1+1}_{r_1-s_1\k}
=\sum_{r_3,s_3\in\bZ}f_{r_3,s_3}(j_3,j_2,j_1)\d^{2j_3+1}_{r_3-s_3\k}=0.
}
}
Proof is given by using the following Lemma-I and Lemma-II.

\noindent{\underbar{\bf Lemma I}}.\refs{\Rmali}~{\it
For \ $\k =k+2 \in { \bf C}\setminus \{ 0 \}$
and the highest weight $j$, parametrized as $ 2j_{r,s}+1=r-s\k $
with $r,s \in {\bf Z}$, such that
{\rm (i) } $r>0$ and $s \geq 0$ or
{\rm (ii)} $r<0$ and $s<0$ ,
there exists a unique null vector $\|\chi_{r,s}\> \in V_j$
of grade $N=rs$ and charge $Q=-r$.
And the null vector in $V_{ j_{r,s}} $ is as follows,
\eqn\?{\eqalign{
&\|\chi_{r,s}\> =
(f_1)^{r+s\k} (f_0)^{r+(s-1)\k} \cdots \cdots
(f_0)^{r-(s-1)\k}(f_1)^{r-s\k}
\|j_{r,s}\>,
\cr
&\|\chi_{r,s}\> =
(f_0)^{-r-(s+1)\k} (f_1)^{-r-(s+2)\k} \cdots \cdots
(f_1)^{-r+(s+2)\k}(f_0)^{-r+(s+1)\k}
\|j_{r,s}\>,
}}
for the cases {\rm(i)}  and {\rm(ii)}  respectively.
}

\noindent{\underbar{\bf Lemma II}}.~{\it
For the null vectors $\|\chi_{r,s}\>\in V_{j_1}$
and $\<\chi_{r,s}\|\in V_{j_3}^*$, we have
\eqn\?{\eqalign{
\<j_3     \|\Phi_{j_2}(z,x)\|\chi_{r,s}\>
&= 
f_{r,s}(j_1,j_2,j_3) z^{ h_3-h_1-rs } x^{ j_1+j_2-j_3-r },
\cr
\<\chi_{r,s}\|\Phi_{j_2}(z,x)\|j_1      \>
&= 
f_{r,s}(j_3,j_2,j_1) z^{ h_3-h_1+rs } x^{ j_1+j_2-j_3+r },
}}
up to some non-zero multiple factors.
}


\appendix{B}{Calculation of the two point function
                         by the $q$-OPE}


\noindent{\bf \S$\,$B.1.}~
We calculate the image $\Phi_{j_2,m_2}(z)\|j_1\>\in V_{j_3}$
of the highest weight vector $\|j_1\>$.
Let
\eqn\?{
\Phi_{j_2,m_2}(z)\|j_1\>
=\d^{j_3-j_1+Q}_{m_2} \sum_{N} z^{ h_3-h_1+N }
q^{ k N + 2j_1 Q }\|N,Q\>_{j_3},
}
where $\|{N,Q}\>_{j_3}$ is the homogeneous components
of grade $N$ and charge $-Q$, such that
\eqn\?{
\|N,Q\>_{j_3}
=\sum_{(\alpha_1,\cdots,\alpha_n)} \b _{\alpha_1,\cdots,\alpha_n}
f_{\alpha_1 \cdots \alpha_n}  \|j_3\>,
}
$N=\sum_i \bar\a_i $, $Q=\sum_i (\a_i - \bar\a_i )$, $\a_i=0,1$
, $\bar\a =1-\a$ and
$f_{\alpha_1 \cdots \alpha_n}= f_{\alpha_1} \cdots f_{\alpha_n}$ .
{}From the definition of the $q$-vertex operator {\Eqvo},
we have the descent equations for $\|{N,Q}\>_{j_3}$ \refs{\RayII}
\eqn\?{\eqalign{
e_1\|N,Q\>_{j_3} &= [-j_1+j_2+j_3-Q+1]\,\|N  ,Q-1\>_{j_3},\cr
e_0\|N,Q\>_{j_3} &= [ j_1+j_2-j_3+Q+1]\,\|N-1,Q+1\>_{j_3}.
}}
{}From these descent equations,
we can calculate the expansion coefficients
$\b _{\alpha_1,\cdots,\alpha_n} $ .


\noindent\underbar{Example}.~
The $q$-OPE of spin $1/2$ {\VO} is
\eqn\?{\eqalign{
\Phi_{\ha ,}{}^0_0(z)\|j\>&=z^{h^+ -h}
(1+q^kz(\b^+_{01}f_{01}+\b^+_{10}f_{10})+\iO(z^2))\|j+{1\over 2}\>,\cr
\Phi_{\ha ,}{}^0_1(z)\|j\>&=z^{h^+ -h} q^{2j}
(\b^+_1f_1+q^kz(\b^+_{011}f_{011}+\b^+_{101}f_{101}+\b^+_{110}f_{110})
                            +\iO (z^2))\|j+{1\over 2}\>,\cr
\Phi_{\ha ,}{}^1_1(z)\|j\>&=z^{h^- -h}
(1+q^kz(\b^-_{10}f_{10}+\b^-_{01}f_{01})+\iO(z^2))\|j-{1\over 2}\>,\cr
\Phi_{\ha ,}{}^1_0(z)\|j\>&=z^{1+h^- -h } q^{k-2j}
(\b^-_0f_0+q^kz(\b^-_{100}f_{100}+\b^-_{010}f_{010}+\b^-_{001}f_{001})
                           +\iO (z^2))\|j-{1\over 2}\>,\cr
}}
where
$h= h_j $, $h^\pm = h_{j\pm\ha}$,
\eqn\?{\eqalign{
\b^+_{  1}&={1     \over [2j+1]        },\cr
\b^+_{ 01}&={[2j+3]\over [2][k+2][2j+1]},\cr
\b^+_{ 10}&={- 1   \over [2][k+2]      },\cr
}\qquad\eqalign{
\b^+_{011}&={-[k+2][2j+2]          \over [k+2j+3][2][k+2][2j+1]},\cr
\b^+_{101}&={ [k+2][2j]+[k+3][2j+3]\over [k+2j+3][2][k+2][2j+1]},\cr
\b^+_{110}&={-[k+3]                \over [k+2j+3][2][k+2]      },\cr
}}
and
\eqn\?{
\b^-_{\alpha_1,\cdots,\alpha_n}(2j)=
\b^+_{\bar\alpha_1,\cdots,\bar\alpha_n}(k-2j).
}


\noindent{\bf \S$\,$B.2.}~
By using the $q$-OPE and one point function,
we can in principle derive the arbitrary $N$ point functions.


\noindent\underbar{Example}.~
 The two point functions for the spin
$\ha$ and arbitrary spin $j_2$ {\VOs} are
\eqn\?{\eqalign{
&\opeOO =z_3^{h_4-h_4^-}z_2^{h_4^- -h_1}\cr
&\,\qquad \times\{1+ {q^{k+2} x \over [2][k+2][k-2j_4+1]}
\Pmat{ [k-2j_4+3][M  ][2j_2-M+1] \cr
      -[k-2j_4+1][M+1][2j_2-M]}+\iO (x^2)\},\cr
\apar
&\opeOI =-z_3^{h_4-h_4^-}z_2^{h_4^- -h_1} x
q^{k-2j_1+1}{[M]\over[k-2j_4+1]}\cr
&\times\{1+ {q^{k+2} x \over [2][k+2][2k-2j_4+3]}
\Pmat{- [k+2][k-2j_4+2][M-1][2j_2-M+2] \cr
      + [k+2][k-2j_4  ][M  ][2j_2-M+1] \cr
      + [k+3][k-2j_4+3][M  ][2j_2-M+1] \cr
      - [k+3][k-2j_4+1][M+1][2j_2-M  ] \cr} + \iO (x^2) \},\cr
}}
\eqn\?{\eqalign{
&\opeIO =-z_3^{h_4-h_4^+}z_2^{h_4^+ -h_1}
q^{2j_1+1}{[2j_2-M+1]\over[2j_4+1]}\cr
&\,\quad\times\{1+ {q^{k+2} x \over [2][k+2][k+2j_4+3]}
\Pmat{- [k+2][2j_4+2][M+1][2j_2-M  ] \cr
      + [k+2][2j_4  ][M  ][2j_2-M+1] \cr
      + [k+3][2j_4+3][M  ][2j_2-M+1] \cr
      - [k+3][2j_4+1][M-1][2j_2-M+2] \cr} +\iO (x^2) \},\cr
\apar
&\opeII =z_3^{h_4-h_4^+}z_2^{h_4^+ -h_1}\cr
&\,\qquad \times\{1+ {q^{k+2} x \over [2][k+2][2j_4+1]}
\Pmat{ [2j_4+3][M  ][2j_2-M+1] \cr
      -[2j_4+1][M-1][2j_2-M+2]}+\iO (x^2) \},\cr
}}
%
and
\eqn\?{\eqalign{
&\opeOOr =z_2^{h_4-h_1^+}z_3^{h_1^+-h_1}\cr
&\,\qquad
\times\{1+ {q^{k+2} x^{-1} \over [2][k+2][2j_1+1]}
\Pmat{ [2j_1+3][2j_2-M+1][M  ] \cr
      -[2j_1+1][2j_2-M  ][M+1] }+\iO ({1 \over x^2})\},\cr
\apar
&\opeOIr =-z_2^{h_4-h_1^+}z_3^{h_1^+ -h_1}q^{2j_4+1}{[M]\over[2j_1+1]}\cr
&\times\{1+ {q^{k+2} x^{-1} \over [2][k+2][k+2j_1+3]}
\Pmat{- [k+2][2j_1+2][2j_2-M+2][M-1] \cr
      + [k+2][2j_1  ][2j_2-M+1][M  ] \cr
      + [k+3][2j_1+3][2j_2-M+1][M  ] \cr
      - [k+3][2j_1+1][2j_2-M  ][M+1] \cr}+\iO ({1 \over x^2}) \},\cr
}}
\eqn\?{\eqalign{
&\opeIOr =-z_2^{h_4-h_1^-}z_3^{h_1^- -h_1} x^{-1}
q^{k-2j_4+1}{[2j_2-M+1]\over[k-2j_1+1]}\cr
&\times\{1+ {q^{k+2} x^{-1} \over [2][k+2][2k-2j_1+3]}
\Pmat{- [k+2][k-2j_1+2][2j_2-M  ][M+1] \cr
      + [k+2][k-2j_1  ][2j_2-M+1][M  ] \cr
      + [k+3][k-2j_1+3][2j_2-M+1][M  ] \cr
      - [k+3][k-2j_1+1][2j_2-M+2][M-1] \cr}+\iO ({1\over x^2})\},\cr
\apar
&\opeIIr =z_2^{h_4-h_1^-}z_3^{h_1^- -h_1}\cr
&\,\qquad \times\{1+ {q^{k+2} x^{-1} \over [2][k+2][k-2j_1+1]}
\Pmat{ [k-2j_1+3][2j_2-M+1][M] \cr
      -[k-2j_1+1][2j_2-M+2][M-1]}+\iO ({1 \over x^2})\},\cr}
}
where $x=z_2/z_3$,
$h_n^{\pm}=h_{j_n\pm\ha}$
and $M=j_1+j_2-j_4+\ha$.


\appendix{C}{The $R$-matrix}


For the finite dimensional representation $V_j(z)$ in \S 2.2,
let $\|M_1,M_2\>=\|1/2,M_1\>\otimes \|j,M_2\>\in V_{1/2}(z) \otimes V_{j}(w)$.
The $R$-matrix, $R(z/w) : V_{1/2}(z) \otimes V_{j}(w)  \->
V_{1/2}(z) \otimes V_{j}(w) $,
defined by $R(z/w) \D'_{z,w} = \D_{z,w} R(z/w)$,
is a $2 \times 2$ block diagonal form in each sector
$\bC \|0, M\> \oplus \bC\|1, M-1\>$,
with $M\in \{1,\cdots,2j-1\}$.

The $R$-matrix $\tR(x)$, which is normalized as $\tR(x)\|0,0\>=\|0,0\>$,
is given explicitly as follows
\eqn\?{
 \Pmat{\tR(x)\|0,M\> \cr \tR(x)\|1,M-1\> \cr}
=\Pmat{a_M(x) & b_M(x) \cr c_M(x) & d_M(x) \cr }
 \Pmat{\|0,M\> \cr \|1,M-1\> \cr },
}
where
\eqn\?{
\Pmat{a_M(x) & b_M(x) \cr
      c_M(x) & d_M(x) \cr }
={1 \over x-q^{M+{\M}}}
\Pmat{ xq^{M }-q^{\M}        &  q^{M }(q^{-{\M}}-q^{\M}) \cr
       xq^{\M}(q^{-M}-q^{M}) & xq^{\M}-q^{M}             \cr},
}
with $\M=2j+1-M$.
They obey the following crossing relation
\eqn\?{\eqalign{
(((\tR(x)^{-1})^{t_1})^{-1})^{t_1} &=
{ (1-x q^{-2j+3})(1-x q^{2j+1}) \over
(1-x q^{-2j+1})(1-x q^{2j+3}) } K^{-1} \tR(x q^{-4}) K,\cr
K &= \Pmat{ q & \cr & q^{-1} \cr } \otimes 1 ,\cr
}}
where $t_1$ means a transpose on the first component.

The $R$-matrix $R(x)$ which is the image of the universal
$R$-matrix can be determined by the crossing relation \refs{\Rfr}
\eqn\?{
(((R(x)^{-1})^{t_1})^{-1})^{t_1} = K^{-1} R (x q^{-4}) K .
}
The relation between $R$ and $\tR$ is then given as
\eqn\?{
 R(x)=f(x) \tR(x), \qquad
f(x)
={(1-xq^{2j_2+3})(1-xq^{-2j_2+1})\over (1-xq^{2j_2+1})(1-xq^{-2j_2+3})}
f(xq^4).
}
The solution for $f(x)$ in the region $x \ll 1$ is
\eqn\?{
f(x)=\prod_{n=0}^{\infty}
{(1-x q^{2j+3+4n})(1-x q^{-2j+1+4n})
\over (1-x q^{2j+1+4n})(1-x q^{-2j+3+4n}) }.
}
This solution can be analytically continued to the whole $x\in\bC$ uniquely.


\listrefs

\bye